\documentclass[twoside,fleqn]{article}
\usepackage{mycrc}

\usepackage{graphicx}

\title{Dynamical Triangulation with Fluctuating Topology}
\author{Bas V.~de Bakker%
  \address{Institute for Theoretical Physics, University of
    Amsterdam,\\ Valckenierstraat 65, 1018 XE Amsterdam, the
    Netherlands}}

\begin{document}

\begin{abstract}
  We consider a dynamical triangulation model of euclidean quantum
  gravity where the topology is not fixed.  This model is equivalent
  to a tensor generalization of the matrix model of two dimensional
  quantum gravity.  A set of moves is given that allows Monte Carlo
  simulation of this model.  Some preliminary results are presented
  for the case of four dimensions.
\end{abstract}

\maketitle

% Kludge to get preprint number on first page.
\def\rightmark{\protect\parbox[t]{2in}{ITFA-94-32\protect\\ November
    1994}}

\section{Introduction}

In the dynamical triangulation model of quantum gravity one usually
considers systems where the spacetime topology is fixed (although the
topology of space is allowed to change with time).  In a path integral
formulation of quantum gravity, summing over the topologies seems a
natural thing to do.

Because the typical curvature fluctuations become larger at smaller
scale, allowing an arbitrary topology will result in a very
complicated structure at the Planck scale \cite{Wh64}.  Such a
spacetime which is full of holes is commonly called spacetime foam.
It was estimated by Hawking \cite{Ha78} that the dominant contribution
to the path integral would come from spacetimes where the Euler
characteristic $\chi$ is of the order of the volume of the spacetime
in Planck units.

\section{Definition of the model}

The partition function of the model in $d$ dimensions is
\begin{equation}
  Z(\kappa, N_d) = \sum_{{\cal T}(N_d)} \exp(\kappa N_{d-2}).
\end{equation}
This expression is the same as for a fixed topology, but here the sum
is over all possible ways to glue a fixed number $N_d$ of
$d$-simplices together, maintaining orientation (i.e.\ only identify
$d-1$ dimensional faces with opposite orientation).  Because the faces
have to be glued in pairs, the number of simplices $N_d$ must be even
in odd dimensions.

At this stage we include disconnected configurations.  Because the
action is a sum of the actions of the connected components, the
Boltzmann weight factorizes, which means that the local physics will
not change.  However, including these configurations will tell us
something about the chance to obtain a particular size of connected
component.  A universe which is most likely to be split up into many
small parts seems an unlikely candidate for the real world.

Naively gluing together simplices will not result in manifolds, but
only in pseudomanifolds.  Several ways to deal with this problem are
conceivable.  First, as nobody knows what spacetime looks like at the
Planck scale, one could argue that this is not a problem.  Second, one
can (at least for $d \leq 4$) locally deform the resulting
pseudomanifold to turn it back into a manifold by removing a small
region around the singular points and pasting in a regular region.
This can be done while changing the total curvature only by an
arbitrarily small amount.

Third, these configurations might be unimportant in the limit $\kappa
\to \infty$ (which will have to be taken, see below).  E.g.\ in three
dimensions for each fixed $N_0$ (and volume $N_3$) the number of edges
(which couples to $\kappa$) is maximal if and only if the
configuration is a real (i.e.\ non-pseudo) manifold.  See
\cite{AmDuJo91} for an explanation.  This is less clear in four
dimensions, though, because a similar reasoning only results in a
closed neighbourhood of a point being bounded by a simplicial manifold
(as opposed to a pseudomanifold), but not necessarily a sphere.

At low $\kappa$ the connected configurations will contribute most as
they have the highest entropy.  At high $\kappa$ the disconnected
configuration will contribute most as they have the lowest action.  It
is a priori not clear whether this change occurs gradually or whether
there is a phase transition.  Suppose for a moment that there is a
sharp crossover at some $\kappa^c$ depending on the volume.  Because
the number of connected configurations rises faster with the volume
than the number of disconnected ones, the value of $\kappa^c$ will
increase logarithmically with the number of simplices.  This means
that in a possible continuum limit, the value of $\kappa$ will have to
be taken to infinity.

\section{Tensor model}

We can formally write down a tensor model which is a generalization of
the well known one matrix model of two dimensional quantum gravity
(see e.g.\ \cite{GiMo92} and references therein).  The partition
function of this tensor model is written in terms of a $k$-dimensional
tensor $M$ of rank $d$.  The tensor $M$ is invariant under an even
permutation of its indices and goes to its complex conjugate under an
odd permutation of its indices.  The partition function for three
dimensions is
\begin{eqnarray}
  Z & = & \int dM^{}_{abc} \exp \left( -\frac{1}{6} M^{}_{abc}
  M^{*}_{abc} + {} \right. \\ & & \left. g M^{}_{abc} M^{*}_{ade}
  M^{}_{bdf} M^{*}_{cef} \vphantom{\frac{1}{1}} \right),
  \label{threedz}
\end{eqnarray}
and for four dimensions it is
\begin{eqnarray}
  Z & = & \int dM^{}_{abcd} \exp \left( -\frac{1}{24} M^{}_{abcd}
  M^{*}_{abcd} + {} \right. \\ & & \left. g M^{}_{abcd} M^{*}_{aefg}
  M^{}_{behi} M^{*}_{cfhj} M^{}_{dgij} \vphantom{\frac{1}{1}} \right)
  .
  \label{fourdz}
\end{eqnarray}
Using the properties of $M$, the action can easily be seen to be real.
The generalization to more dimensions should be obvious from these
expressions.

These expressions are only formal, because the interaction term can be
negative for any $g$ and is of higher order than the gaussian term.
Therefore, these integrals will not converge.  However, if we expand
these expressions in $g$, each term of the expansion is well defined
and the dual of each of its Feynman diagrams is a DT configuration of
size $N_d$ equal to the order of $g$ used.  Each propagator carries
two sets of $d$ indices.  Each vertex of the tensor model corresponds
to a $d$-simplex, and each propagator between them corresponds to the
identification of two $d-1$ dimensional faces of simplices.  The sets
of indices at each end of a propagator must be an odd permutation of
each other, making sure that the simplicial complex has an
orientation.  The contribution of a particular diagram is
\begin{equation}
  g^{N_d} k^{N_{d-2}} = \exp ( \ln(g) N_d + \ln(k) N_{d-2} ),
\end{equation}
which is precisely proportional to the Boltzmann weight of the DT
configuration according to the Regge-Einstein action, with the
identification $\kappa = \ln k$.

A similar model using hypercubes was introduced in \cite{We82}.  The
model described above has been discussed for three dimensions in
\cite{AmDuJo91}.  A different generalization of the matrix model where
the dimension of the matrix couples to the number of points in the DT
configuration (which means one does not get the Regge-Einstein action
in more that 3 dimensions) has been discussed in \cite{Gr91}.

\section{Monte Carlo simulation}

A move can most easily be described in the tensor model formulation.
It consists of first cutting two propagators and then randomly
reconnecting them.  Due to the orientability, there are $d!/2$ ways to
connect two propagators.  In the dynamical triangulation model, this
corresponds to cutting apart the simplices at each side of two of the
$d-1$ dimensional faces and pasting these faces together.

Unlike the case of fixed topology with $(k,l)$ moves, one can easily
see that these moves are ergodic.  The number of moves needed to get
{}from one configuration to another is $O(N_d)$, raising none of the
computability problems associated with the fixed topology
\cite{NaBe93}.  The non-existance of a classification of
four-topologies and their unrecognizability is usually mentioned as a
problem for the summation over topologies.  In dynamical
triangulation, however, it seems to be more a problem for fixing the
topology.

We use the standard Metropolis test to accept or reject the moves.
Because (again unlike the fixed topology case) the number of possible
moves does not depend on the configuration, detailed balance is
easily obtained.

One could restrict the simulation to connected configurations by
checking connectedness for each move accepted by the Metropolis test.
This might be rather slow, because this is not a local test.  Also,
although it seems very plausible, it is not clear whether this would
be ergodic in the space of connected configurations.

\section{Results}

We have simulated this model in four dimensions at a volume $N_4 =
500$.  We used a hot start, that is a configuration with completely
random connections.  This would be an equilibrium configuration at
$\kappa = 0$.

\begin{figure}[t]
\includegraphics[width=75mm]{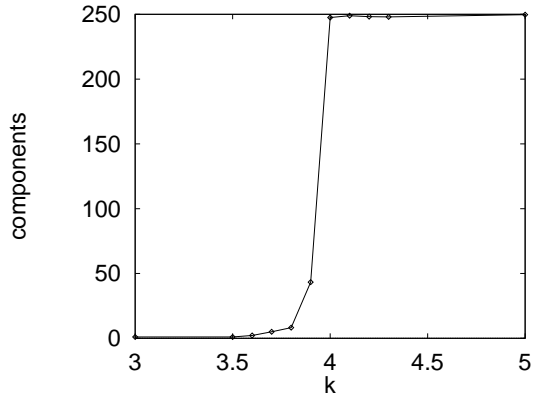}
\caption{The number of connnected components as a function of
  $\kappa$ in the four dimensional model.}
\label{cmpfig}
\end{figure}
The number of connected components is plotted in figure \ref{cmpfig}.
We see that at low $\kappa$ the average number of components is almost
one, while at high $\kappa$ the average number of components is almost
equal to the maximum number of $N_d/2$.  Already at this low number of
simplices, the change between few and many components looks quite
sharp.

Unfortunately, the acceptance rates for these moves are quite low for
$\kappa$ near $\kappa^c$, of the order of 0.1\%.  One of the reasons
for this is that the proposed moves are not local in the sense that
they can connect any two points in the complex.  This means that
simulating much larger systems will probably not be feasible with only
these moves.

\section{Discussion}

The main problem to investigate is the existence of a sensible
continuum limit.  Although, due to the factorially increasing number
of configurations, the grand canonical partition function in the
normal definition does not exist, this does not exclude that the local
behaviour of the system might show scaling for large volumes.

In two dimensions a limit of the grandcanonical partition function,
the so-called double scaling limit, is known
\cite{DblScl}.  In this limit we also see that the
coupling $\kappa$ has to be taken to infinity.

\section*{Acknowledgement}

The author would like to thank Jan Smit for discussions.

% End of preprint number kludge
\def\rightmark{}

\end{document}